\documentclass[traditabstract]{aa}

\usepackage{graphicx}
\usepackage{natbib}
\usepackage[dvipdfm]{hyperref}
\usepackage{amssymb}
\usepackage{bm}

\newcommand {\eg} {{\it e.g.}}

\newcommand {\be} {\begin{equation}}
\newcommand {\ee} {\end{equation}}
\newcommand {\bea} {\begin{eqnarray}}
\newcommand {\eea} {\end{eqnarray}}

\defcitealias{2009MNRAS.395..524N}{Paper I}

\begin{document}

\title{Polarization of synchrotron emission from relativistic reconfinement shocks with ordered magnetic fields}
\titlerunning{Polarization from reconfinement shocks with ordered magnetic fields}

\author{Krzysztof Nalewajko\inst{1,2}\thanks{\email{knalew@colorado.edu}} \and Marek Sikora\inst{2}}
\authorrunning{Nalewajko \and Sikora}

\institute{University of Colorado, 440 UCB, Boulder, CO 80309, USA
\and
Nicolaus Copernicus Astronomical Centre, Bartycka 18, 00-716 Warsaw, Poland}

\abstract{We calculate the polarization of synchrotron radiation produced at the relativistic reconfinement shocks, taking into account globally ordered magnetic field components, in particular toroidal and helical fields. In these shocks, toroidal fields produce high parallel polarization (electric vectors parallel to the projected jet axis), while chaotic fields generate moderate perpendicular polarization. Helical fields result in a non-axisymmetric distribution of the total and polarized brightness. For a diverging downstream velocity field, the Stokes parameter $U$ does not vanish and the average polarization is neither strictly parallel nor perpendicular. A distance at which the downstream flow is changing from diverging to converging can be easily identified on polarization maps as the turning point, at which polarization vectors switch, e.g., from clockwise to counterclockwise.}


\maketitle

\section{Introduction}
\label{sec_intro}

Relativistic jets are present in all radio-loud active galaxies and are responsible for the bulk of their non-thermal emission. The spectral energy distributions of these sources usually show two broad components. The low-energy one, extending from radio wavelengths to the optical/UV band, and in some cases even to hard X-rays, is commonly interpreted as the synchrotron radiation. This interpretation is supported by significant linear polarization measured routinely in the optical \citep[\eg,][]{1978ApJ...220L..67S,1991ApJ...375...46I,1992ApJ...398..454W,2011PASJ...63..639I}, radio \citep[\eg,][]{2001ApJ...562..208L,2002ApJ...577...85M,2002ApJ...568...99H,2003ApJ...589..733P,2011MNRAS.412..318M} and other bands.

Polarization measurements in sources dominated by relativistic jets provide important constraints on the structure of the dominant emitting regions. In particular, a significant polarization degree indicates an anisotropic distribution of magnetic fields. This anisotropy may reflect a large-scale order in the magnetic field lines \citep{1984RvMP...56..255B} or may arise from chaotic magnetic fields compressed at shocks \citep{1980MNRAS.193..439L,1985ApJ...298..301H,1988ApJ...332..678J} or sheared at the jet boundary layer \citep{1981ApJ...248...87L,1990MNRAS.242..616M}. For purely chaotic magnetic fields compressed by perpendicular or conical shocks, high polarization degrees are found only when the polarization (electric) vectors are parallel to the projected jet axis (``parallel polarization''), while for polarization vectors perpendicular to the jet axis (``perpendicular polarization''), the polarization degrees are below $10\%$ \citep{1990ApJ...350..536C}. To produce higher perpendicular polarization degrees, \cite{2006MNRAS.367..851C} introduced a large-scale poloidal magnetic field component. Diverging conical shocks can be caused by collision of the jet with a dense cloud \citep{1985ApJ...295..358L} or appear when the jet rapidly becomes overpressured with respect to its environment \citep{2012ApJ...752...92A}.

A different shock structure can result if a jet becomes underpressured. Then, 
the interaction between the relativistic jet and its gaseous environment leads to the formation of so-called reconfinement or recollimation shocks \citep{1983ApJ...266...73S,1997MNRAS.288..833K}. These shocks deflect the jet flow and focus it on a cross section much smaller compared to that of a freely propagating jet. \citet[hereafter \citetalias{2009MNRAS.395..524N}]{2009MNRAS.395..524N} calculated the linear polarization of synchrotron emission associated with reconfinement shocks for purely chaotic magnetic fields. He found that perpendicular polarization degrees can exceed $20\%$. Because a reconfinement shock is an ensemble of conical shocks, this result appears to be in conflict with the work of \cite{1990ApJ...350..536C}. The crucial difference is that while \cite{1990ApJ...350..536C} assumed a parallel upstream flow, in \citetalias{2009MNRAS.395..524N} a spherically diverging upstream flow was considered. Because parallel upstream flows are often adopted in studies of relativistic jet polarization \citep{2005MNRAS.360..869L,2006MNRAS.367..851C}, we would like to point out that jet divergence makes a substantial difference in the resulting polarization.

Reconfinement shocks with purely chaotic magnetic fields cannot account for the high parallel polarization often observed in blazars. Therefore, we extend the study presented in \citetalias{2009MNRAS.395..524N} to include large-scale ordered magnetic field components. High parallel polarization can be achieved by introducing a toroidal magnetic field component, but in general ordered magnetic fields may consist of both toroidal and poloidal components, forming a helical structure. In Section \ref{sec_pol}, we describe our model of synchrotron emission and polarization from relativistic reconfinement shocks, introducing a simply parametrized family of global helical magnetic fields. In Section \ref{sec_res}, we present the results demonstrating the effect of mixing chaotic and ordered (toroidal) magnetic fields (Section \ref{sec_pol_tor}), and the effect of changing the pitch angle of the purely helical magnetic field (Section \ref{sec_pol_heli}). Our results are discussed in Section \ref{sec_dis} and summarized in Section \ref{sec_sum}. Preliminary results were already presented in \cite{2010ASPC..427..205N}.

\section{Synchrotron emission and polarization}
\label{sec_pol}

We used the model of the structure of reconfinement shocks from \cite{2009MNRAS.392.1205N} to investigate its polarimetric properties. Our attention is focused on the optical wavelengths, which in blazars are dominated by the synchrotron emission of highly relativistic electrons. The cooling time scale for these particles is very short, so that the emitting region is closely aligned with the sites of particle acceleration.\footnote{Observed frequency of the peak of synchrotron emission produced by electrons of Lorentz factor $\gamma$ is $\nu_{\rm syn}=\mathcal{D}B\gamma^2\;4.8\;{\rm MHz}$. Assuming that cooling is dominated by radiative losses due to the synchrotron emission, the cooling length scale, a typical distance covered by a particle before losing a significant fraction of its energy, is
\begin{displaymath}
l_{\rm cool}'\sim ct_{\rm cool}'\sim 6\pi\frac{m_{\rm e}c^2}{\sigma_{\rm T}}\frac{1}{\gamma B^2}\sim 7.5\;{\rm pc}\times\frac{\mathcal{D}^{1/2}}{B^{3/2}}\left(\frac{4.8\;{\rm MHz}}{\nu_{\rm syn}}\right)^{1/2}\,.
\end{displaymath}
For magnetic fields strength $B=1\;{\rm G}$ and Doppler factor $\mathcal{D}=10$, optical-emitting electrons ($\nu_{\rm syn}=6\times 10^{14}\;{\rm Hz}$) have $l_{\rm cool}'\sim 2\;{\rm mpc}$. For comparison, high-frequency radio-emitting electrons ($\nu_{\rm syn}=43\;{\rm GHz}$) have $l_{\rm cool}'\sim 0.25\;{\rm pc}$.}
Relativistic shock fronts provide natural conditions for efficient particle acceleration \citep[\eg][]{2001MNRAS.328..393A}. We are not concerned with details of these processes, it is only assumed that the relativistic electrons tap a constant fraction $\eta_{\rm rad}$ of the post-shock internal energy. Since radiative cooling of these electrons is very efficient, practically their entire energy is transferred to the non-thermal radiation, which includes the synchrotron and inverse-Compton components. Although the inverse-Compton luminosity exceeds the synchrotron luminosity in radio-loud quasars, here we assume that their ratio is roughly constant along the jet. This is justified because both the magnetic energy density and the total synchrotron and external radiation energy densities scale roughly like $r^{-2}$ \citep{2009ApJ...704...38S}. Hence, a constant fraction $\eta_{\rm syn}$ of the energy contained in relativistic electrons goes to the synchrotron radiation. In this case, the power emitted in the form of synchrotron radiation can be determined from the rate of energy transfer into relativistic electrons, which in turn is proportional to the rate of energy dissipation.

Let $I'(\Omega')$ be the synchrotron radiation intensity, as measured in the co-moving frame of the post-shock fluid (denoted with a prime). It is directly related to the emitted flux density, which is Lorentz-invariant, and can therefore be written in the external frame:
\bea
\label{eq_Itot}
\Phi' &=& \int_{\Omega'}d\Omega'I'(\Omega') = \frac{\Delta E'}{\Delta A'\,\Delta t'} = \frac{\Delta E}{\Delta A\,\Delta t}
=\nonumber\\
&=&
\eta_{\rm rad}\eta_{\rm syn}\frac{\Delta E_{\rm diss}}{\Delta A\,\Delta t}
=\eta_{\rm rad}\eta_{\rm syn}\frac{\Delta E_{\rm kin,j}-\Delta E_{\rm kin,s}}{\Delta A\,\Delta t}\,,
\eea
where $\Delta A$ is the shock front area. Consider a shock front element covering distance $\Delta z$ and azimuthal angle $\Delta\phi$. The surface area of this element is
\be
\Delta A_{\rm j(s)} = \frac{r_{\rm s}\;\Delta z\;\Delta\phi}{\cos\theta_{\rm j(s)}}\,,
\ee
where $r_{\rm s}$ is the local jet radius and $\theta_{\rm j(s)}$ is the angle between the fluid velocity vector and the jet axis. The volume of jet (shocked jet) matter crossing this shock front element over time $\Delta t$ is
\be
\Delta V_{\rm j(s)}
= \beta_{\rm j(s)}c\;\Delta t\;\sin\delta_{\rm j(s)}\;\Delta A_{\rm j(s)}\,,
\ee
where $\beta_{\rm j(s)}$ is the fluid velocity and $\delta_{\rm j(s)}$ is the angle made by the velocity vector to the shock tangent. Its kinetic energy is then
\be
\Delta E_{\rm kin,j(s)}=(\Gamma_{\rm j(s)}-1)\Gamma_{\rm j(s)}\rho_{\rm j(s)}c^2\Delta V_{\rm j(s)}\,,
\ee
where $\Gamma_{\rm j(s)}=(1-\beta_{\rm j(s)})^{-1/2}$ is the corresponding Lorentz factor and $\rho_{\rm j(s)}$ is the co-moving mass density. Substituting this into Equation \ref{eq_Itot}, we can calculate $\Phi'$ for every shock front element.

Radiation produced in the post-shock co-moving frame is subject to relativistic Doppler boosting into the external frame. Since this emitting region is fixed in the external frame, the transformation law for the stationary pattern should be used instead of a moving blob \citep[see Appendix A in][]{1997ApJ...484..108S}.  Let $\bm{k}$ be the unit vector toward the observer and $\bm{e}$ the unit vector along the fluid velocity. Then, we have
\be
\label{eq_pol_boost}
I(\bm{k}) = \frac{\mathcal{D}_{\rm s}^3}{\Gamma_{\rm s}}I'(\bm{k}')\,,
\ee
where
\be
\mathcal{D}_{\rm s}=\frac{1}{\Gamma_{\rm s}\left(1-\beta_{\rm s}\bm{k}\cdot\bm{e}\right)}
\ee
is the Doppler factor. From the law of relativistic aberration, we find
\be
\bm{k}'=\mathcal{D}_{\rm s}\left\{\bm{k}+\left[\left(\Gamma_{\rm s}-1\right)\left(\bm{k}\cdot\bm{e}\right)-\Gamma_{\rm s}\beta_{\rm s}\right]\bm{e}\right\}\,.
\ee

Polarization of the synchrotron emission can be described by the polarization degree $\Pi$ and the positional angle of the electric vector $\chi_{\rm E}$ (EVPA; ``polarization angle''). The maximum polarization degree of the synchrotron radiation, corresponding to a uniform magnetic field, is $\Pi_{\rm max}=(3p+3)/(3p+7)$, where $p$ is the index of the electron energy distribution $N(\gamma)\propto\gamma^{-p}$ \citep[\eg,][]{1980MNRAS.193..439L}. When combining polarized emission from many sources, it is convenient to use the Stokes parameters
\bea
Q &=& \Pi\;\cos{(2\chi_{\rm E})}\;I\,, \\
U &=& \Pi\;\sin{(2\chi_{\rm E})}\;I\,.
\eea
When EVPA is measured from the same reference positional angle for all sources, the Stokes parameters are linearly additive. In this work, $\chi_{\rm E}$ is always measured from the projected direction of the jet axis $z$. To calculate the polarization angle, one needs to know the direction of the magnetic field, represented by a unit vector $\bm{b}$.

Following \citetalias{2009MNRAS.395..524N}, we introduce a cartesian coordinate system $(x,y,z)$, in which the jet axis is aligned with the $z$-axis and the observer is in the $xz$-plane, inclined to the jet axis at angle $\theta_{\rm obs}$, hence
\be
\bm{k}=[\sin\theta_{\rm obs},0,\cos\theta_{\rm obs}]\,.
\ee
The post-shock fluid velocity $\bm\beta_{\rm s}=\beta_{\rm s}\bm{e}$ is inclined to the jet axis at angle $\theta_{\rm s}$, hence
\be
\bm{e}=[\sin\theta_{\rm s}\cos\phi,\sin\theta_{\rm s}\sin\phi,\cos\theta_{\rm s}]\,.
\ee
Next, we introduce an orthogonal coordinate system in the plane of the sky:
\bea
\bm{v}' &=& \left[\frac{k_{\rm z}'}{\sqrt{1-k_{\rm y}'^2}}, 0, \frac{-k_{\rm x}'}{\sqrt{1-k_{\rm y}'^2}}\right]\,, \\
\bm{w}' &=& \left[\frac{-k_{\rm x}'k_{\rm y}'}{\sqrt{1-k_{\rm y}'^2}}, \sqrt{1-k_{\rm y}'^2}, \frac{-k_{\rm y}'k_{\rm z}'}{\sqrt{1-k_{\rm y}'^2}}\right]\,.
\eea
The positional angle of the magnetic field vector is
\be
\chi_{\rm B}'=\arctan\left(\frac{\bm{b}'\cdot\bm{w}'}{\bm{b}'\cdot\bm{v}'}\right)\,.
\ee
From this, we subtract the positional angle of the fluid velocity vector:
\be
\chi_{\rm e}'=\arctan\left(\frac{\bm{e}\cdot\bm{w}'}{\bm{e}\cdot\bm{v}'}\right)\,.
\ee
The relative electric vector polarization angle is
\be
\chi_{\rm E}'=\chi_{\rm B}'-\chi_{\rm e}'-\frac{\pi}{2}\,.
\ee
When transformed back to the external frame, the total flux is boosted according to Equation (\ref{eq_pol_boost}). The polarization degree and the polarization angle, measured with respect to the projection of transformation velocity vector onto the plane of the sky, are Lorentz-invariant. To find the absolute polarization angle in the external frame, one needs to add the positional angle of the fluid velocity with respect to the jet axis, which is given by
\be
\chi_{\rm e} = \arctan\left(\frac{\sin\theta_{\rm s}\sin\phi}{\sin\theta_{\rm s}\cos\theta_{\rm obs}\cos\phi-\cos\theta_{\rm s}\sin\theta_{\rm obs}}\right)\,.
\ee
The final formula for the electric vector polarization angle is
\be
\chi_{\rm E}=\chi_{\rm B}'-\chi_{\rm e}'+\chi_{\rm e}-\frac{\pi}{2}\,.
\ee

Using a particular model of the reconfinement shock structure, one can calculate Stokes parameters for every shock front element and then integrate them to produce simulated two-dimensional maps, longitudinal one-dimensional profiles, or the average signal. In the following sections, we consider different distributions of the magnetic fields and calculate the polarization of light observed at different viewing angles.

\subsection{Chaotic magnetic field}
\label{sec_pol_chaos}

For a chaotic magnetic field of initially isotropic distribution, considered in detail in \citetalias{2009MNRAS.395..524N}, the symmetry is broken by compression at the shock front. The resulting magnetic field distribution can be characterized by the shock compression factor $\kappa=(\rho_{\rm j}/\rho_{\rm s})$ and the unit vector normal to the shock front $\bm{n}$. For a shock front element located at an azimuthal angle $\phi$ and inclined to the jet axis by angle $\alpha_{\rm s}$,
\be
\bm{n}=[\cos\alpha_{\rm s}\cos\phi,\cos\alpha_{\rm s}\sin\phi,-\sin\alpha_{\rm s}]\,.
\ee
This unit vector transforms into the co-moving frame as
\be
\bm{n}'=[\cos\alpha_{\rm s}'\cos\phi,\cos\alpha_{\rm s}'\sin\phi,-\sin\alpha_{\rm s}']\,,
\ee
where $\alpha_{\rm s}'$ is the co-moving inclination angle:
\be
\alpha_{\rm s}'=\theta_{\rm s}-\arctan\left[\Gamma_{\rm s}\tan\left(\theta_{\rm s}-\alpha_{\rm s}\right)\right]\,.
\ee
Polarization degree of the resulting synchrotron radiation is given by (\citealt{1985ApJ...298..301H}; see also Appendix \ref{app_pol})
\be
\Pi(\bm{k}') = \Pi_{\rm max}\frac{\left(1-\kappa^2\right)\left[1-\left(\bm{k}'\cdot\bm{n}'\right)^2\right]}{2-\left(1-\kappa^2\right)\left[1-\left(\bm{k}'\cdot\bm{n}'\right)^2\right]}\,.
\ee
The total intensity is related to the total radiation flux by\footnote{The co-moving anisotropy of the total intensity has been neglected in \cite{2009MNRAS.395..524N}, but the consequences of this effect are not significant.}
\be
I'(\bm{k}') = \left(\frac{3}{8\pi}\right)\left\{\frac{2-\left(1-\kappa^2\right)\left[1-\left(\bm{k}'\cdot\bm{n}'\right)^2\right]}{2+\kappa^2}\right\}\Phi'\,.
\ee
The magnetic polarization vector is tangent to both the shock front and the plane of the sky, therefore $\bm{b}'\propto (\bm{k}'\times\bm{n}')$.

\subsection{Ordered magnetic field}
\label{sec_pol_order}

Models of the formation of relativistic jets involve magnetic fields that are ordered on a global scale (\citealt{1977MNRAS.179..433B,1982MNRAS.199..883B}; for a recent review see \citealt{2010LNP...794..233S,2010PhyU...53.1199B}). Initially of poloidal orientation, at large distances the magnetic fields are shaped by the rotation of the central engine and the jet lateral expansion. Their most general form is that of a tightly wound helical structure, which at large distances is increasingly dominated by the toroidal component. Here, we introduce a one-parameter family of globally ordered magnetic field structures that can be easily implemented into the reconfinement shock models.

At every point $(z,r,\phi)$ of the jet region, the magnetic field, measured in the fluid co-moving frame, has a toroidal component $B_{\rm t}$ along the azimuthal angle $\phi$ and a poloidal component along the fluid velocity vector, i.e. the $R=\sqrt{r^2+z^2}$ coordinate. The third component, which would be directed along the polar angle $\theta=\arctan(r/z)$, is neglected. The combination of these components forms a helix of pitch angle $\alpha_{\rm B}=\arctan(B_{\rm p}/B_{\rm t})$.

For each magnetic field component, a relation between its values across the different locations needs to be established. First of all, the magnetic field structure is assumed to be axisymmetric, i.e. $B_{\rm p}$ and $B_{\rm t}$ are independent of $\phi$. For a particular fluid element, conservation of the magnetic flux implies the following scaling laws: $B_{\rm p}\propto R^{-2}$ and $B_{\rm t}\propto R^{-1}$. However, the dependence of the magnetic field components on the polar angle $\theta$ is arbitrary. Here, a choice is made that for the same values of $R$, $B_{\rm p}$ is independent of $\theta$ and $B_{\rm t}\propto\sin\theta$. This scaling satisfies a condition that $B_{\rm t}$ has to vanish at $\theta=0$. The explicit formulas for the magnetic field components at any point within the jet region are
\be
B_{\rm p}=\frac{C_{\rm p}}{R^2}\,,
\qquad
B_{\rm t}=\frac{C_{\rm t}\sin\theta}{R}\,.
\ee
We relate the constants $C_{\rm p}$ and $C_{\rm t}$ to the value of the pitch angle $\alpha_{\rm B,m}$ at the jet maximum radius $r_{\rm m}$ at $R_{\rm m}=\sqrt{r_{\rm m}^2+z_{\rm m}^2}$.
\be
\tan\alpha_{\rm B,m}=\frac{C_{\rm p}}{C_{\rm t}r_{\rm m}}\,.
\ee
Thus, the pitch angle at any point in the jet region is
\be
\label{eq_pol_alphab}
\tan\alpha_{\rm B}=\left(\frac{r_{\rm m}}{r}\right)\tan\alpha_{\rm B,m}\,.
\ee
Since $r<r_{\rm m}$, $\alpha_{\rm B,m}$ is the minimum value of the pitch angle that can be achieved anywhere at the shock front. Eliminating $C_{\rm p}$, we calculate the magnetic field components before crossing the shock front
\be
B_{\rm p,j}=C_{\rm t}\frac{r_{\rm m}\tan\alpha_{\rm B,m}}{R_{\rm s}^2}\,,
\qquad
B_{\rm t,j}=C_{\rm t}\frac{\sin\theta_{\rm j}}{R_{\rm s}}\,.
\ee
The value of $C_{\rm t}$ is not important for the resulting polarization, hence the only free parameter of the magnetic field structure is $\alpha_{\rm B,m}$. Across the shock front, the poloidal component is conserved, $B_{\rm p,s}=B_{\rm p,j}$, while the toroidal component is compressed by a factor $\kappa=(\rho_{\rm j}/\rho_{\rm s})$, i.e. $B_{\rm t,s}=B_{\rm t,j}/\kappa$. Thus, the post-shock pitch angle is
\be
\label{eq_pol_alphabs}
\tan\alpha_{\rm B,s} = \frac{B_{\rm p,s}}{B_{\rm t,s}} = \left(\frac{\kappa\,r_{\rm m}}{r_{\rm s}}\right)\tan\alpha_{\rm B,m}\,.
\ee
In the coordinate system introduced in Section \ref{sec_pol} to calculate the Stokes parameters, the magnetic field direction is
\be
\label{eq_pol_vecb_heli}
\bm{b}' \propto B_{\rm p,s}\bm{e}+B_{\rm t,s}\bm{\phi}\,,
\ee
where $\bm{e}$ is the unit vector along the fluid velocity and $\bm\phi=[-\sin\phi,\cos\phi,0]$. Since the magnetic field is locally uniform, the polarization degree is equal to $\Pi_{\rm max}$ and the total intensity is
\be
I'(\bm{k}')=\left(\frac{3}{8\pi}\right)\left[1-\left(\bm{b}'\cdot\bm{k}'\right)^2\right]\Phi'
\ee
(see Appendix \ref{app_pol}).

\section{Results}
\label{sec_res}

In this section, we extend the calculations presented in \citetalias{2009MNRAS.395..524N} to cases involving ordered magnetic field components. We used the same model of the reconfinement shock structure as in \citetalias{2009MNRAS.395..524N}. It is assumed that the jet plasma is cold and lowly magnetized, so that the jet does not accelerate and expands conically, and the shock equations can be solved in hydrodynamical approximation. In all models presented here, a jet with the Lorentz factor of $\Gamma_{\rm j}=10$ and the half-opening angle of $\Theta_{\rm j}=5^\circ$ interacts with the external medium of pressure distribution $p_{\rm e}\propto r^{-\eta}$ with $\eta=0$. We do not explore the dependence of the results on the values of these parameters, but rather on the viewing angle $\theta_{\rm obs}$, and on the parameters describing the magnetic field structure.

\subsection{Toroidal magnetic field}
\label{sec_pol_tor}

For $\alpha_{\rm B,m}=0$, in our model of the global magnetic field structure, the poloidal component vanishes everywhere, leaving the purely toroidal field. In this section, we investigate the polarimetric properties of reconfinement shocks filled with a toroidal field component or a combination of chaotic and toroidal components. This combination is calculated by introducing a parameter $f$ that describes what fractions of the total intensity are emitted from each component:
\be
I'=f I_{\rm ordered}'+(1-f)I_{\rm chaotic}'\,.
\ee
It can also be understood as the fraction of the total magnetic energy density $u_{\rm B}=B^2/(8\pi)$ contained in the ordered magnetic field component. Polarized intensities $Q'$ and $U'$ are combined in the same way.

\begin{figure}
\includegraphics[width=\columnwidth]{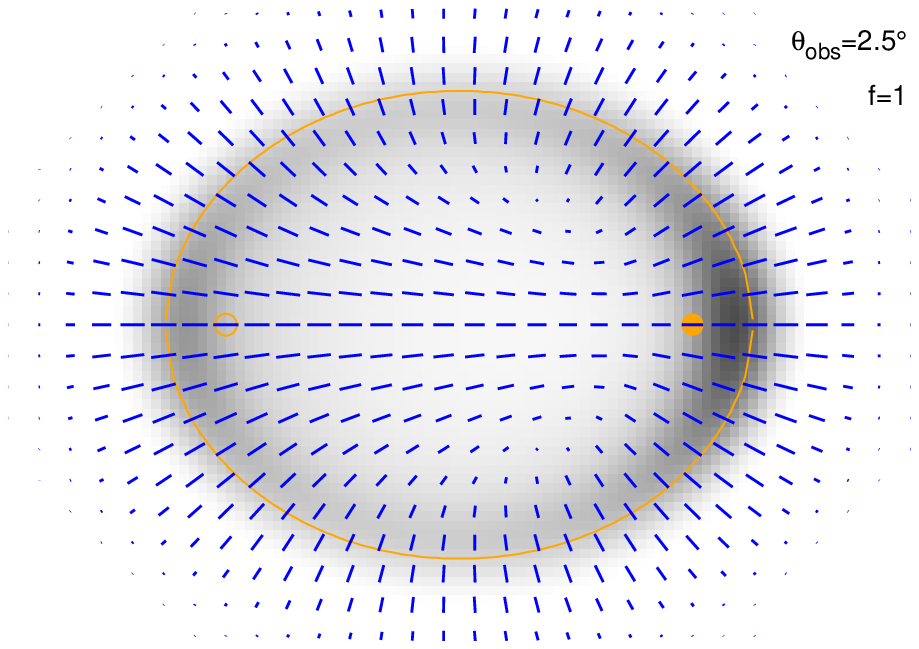}
\includegraphics[width=\columnwidth]{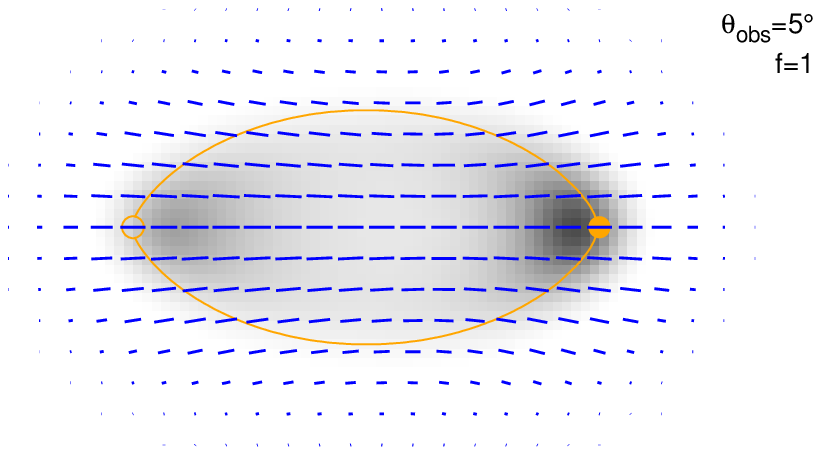}
\caption{Synthetic emission maps of the reconfinement shock with jet Lorentz factor $\Gamma_{\rm j}=10$, opening angle $\Theta_{\rm j}=5^\circ$ and external pressure index $\eta=0$, filled with a toroidal magnetic field and seen from different viewing angles $\theta_{\rm obs}$. The jet origin is marked with an \emph{empty circle}, the reconfinement point with a \emph{filled circle}. The outline of the shock fronts is shown with \emph{orange lines}. \emph{Gray shading} indicates the intensity of synchrotron radiation, while the \emph{blue bars} mark the electric polarization vector of length proportional to the polarization degree.}
\label{fig_pol_map-1a-f100}
\end{figure}

In Figure \ref{fig_pol_map-1a-f100}, we present emission maps of reconfinement shocks filled with toroidal magnetic fields. They can be directly compared to the maps calculated for a chaotic magnetic field, presented in \citetalias{2009MNRAS.395..524N}, since the shock structure is the same. The total flux distribution is more concentrated toward the jet axis. Shock outlines at the largest jet radius are less pronounced for $\theta_{\rm obs}=\Theta_{\rm j}/2$ and barely visible for $\theta_{\rm obs}=\Theta_{\rm j}$. The longitudinal profiles of the total flux calculated for toroidal magnetic fields (not shown) are similar to those for the chaotic magnetic field. A major difference is that for $\theta_{\rm obs}\le\Theta_{\rm j}$, the central areas between the two emission peaks are up to $\sim 30\%$ dimmer. These differences arise because emission from a uniform magnetic field is more anisotropic than emission from a two-dimensional chaotic distribution. In the areas close to the shock outline at the maximum radius, the magnetic field lines are pointing more closely toward the observer, and consequently less emission can be seen.

The polarization maps are very different from those for chaotic magnetic fields. For $\theta_{\rm obs}=\Theta_{\rm j}$, we find a very simple structure of parallel polarization vectors and polarization degrees very close to the maximum, as is generally expected for toroidal magnetic fields. However, the picture is more complicated for a smaller viewing angle. For $\theta_{\rm obs}=\Theta_{\rm j}/2$, in the vicinity of the shock outline the polarization vectors are perpendicular to the projected shock fronts, which is very similar to the case of chaotic fields. But the central regions of the image show strong parallel polarization. Because the total flux is dominated by the areas close to the projected jet axis, their parallel polarization is expected to dominate the perpendicularly polarized areas.

\begin{figure}
\includegraphics[width=\columnwidth]{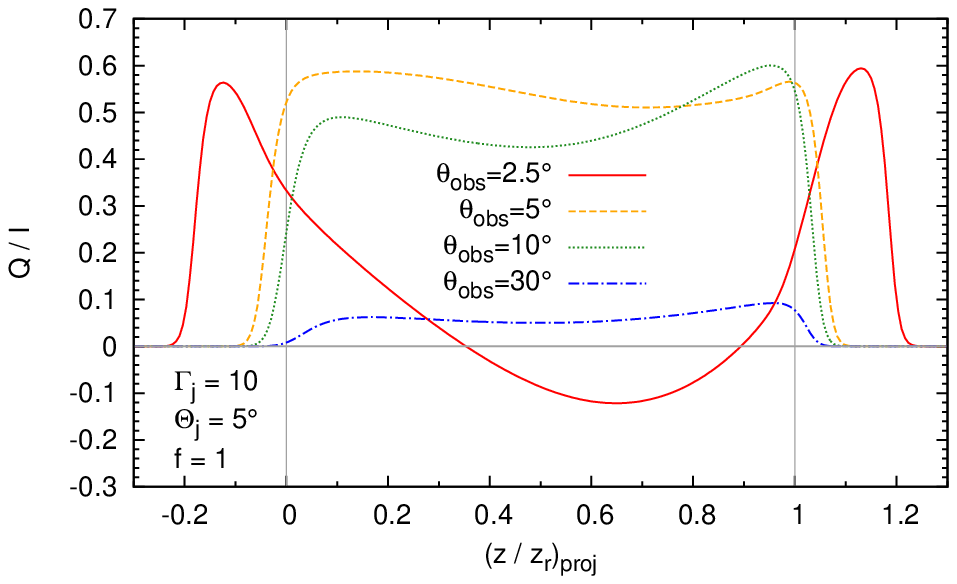}
\includegraphics[width=\columnwidth]{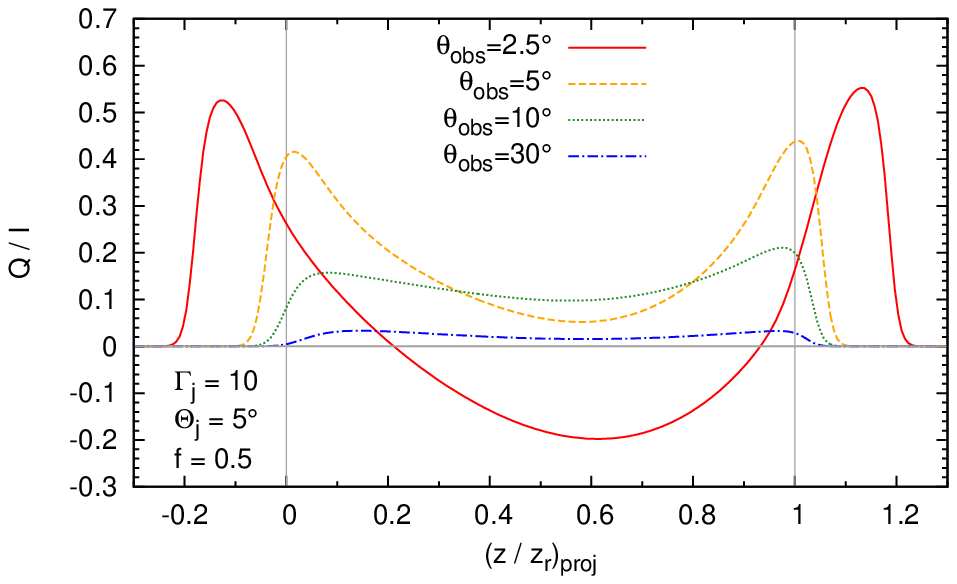}
\caption{Profiles of the polarization degree along the projected jet axis calculated for different viewing angles $\theta_{\rm obs}$ in two cases: a pure toroidal magnetic field (\emph{upper panel}; same as models shown in Figure \ref{fig_pol_map-1a-f100}) and an even mix of toroidal and compressed magnetic fields (\emph{lower panel}). Projected coordinate $(z/z_{\rm r})_{\rm proj}$ equals $0$ for the jet origin and $1$ for the reconfinement point.}
\label{fig_pol_prof-pol-1a-tor}
\end{figure}


In Figure \ref{fig_pol_prof-pol-1a-tor}, we show the longitudinal profiles of polarization degree for several viewing angles for a toroidal magnetic field. Polarization is generally parallel to the jet axis, with the exception of the middle region for $\theta_{\rm obs}=\Theta_{\rm j}/2$, where the maximum polarization degree is $\sim 12\%$. Parallel polarization degrees reach very high values, almost $\sim 60\%$, for $\theta\le 2\Theta_{\rm j}$. For $\theta_{\rm obs}=30^\circ$, the polarization degree reaches only $\sim 9\%$. The polarization degree profile is very uniform along the jet axis for $\theta_{\rm obs}\ge\Theta_{\rm j}$, while for $\theta_{\rm obs}=\Theta_{\rm j}/2$ it is qualitatively similar to the profile calculated for the chaotic magnetic field.

In the lower panel of Figure \ref{fig_pol_prof-pol-1a-tor}, we show the polarization degree profiles for an even mix of toroidal and compressed magnetic fields ($f=0.5$). Since the total flux ($I$) profiles for both magnetic field distributions are similar, the combined polarization degree closely reflects the linear combination used to calculate the combined polarized flux $Q$. Indeed, the polarization profiles for $f=0.5$ have properties on average between cases $f=0$ (\citetalias{2009MNRAS.395..524N}) and $f=1$ (upper panel of Figure \ref{fig_pol_prof-pol-1a-tor}). For $\theta_{\rm obs}=\Theta_{\rm j}/2$, the profile is similar to that for $f=1$, with slightly lower parallel polarization maxima and higher perpendicular polarization maximum. For higher viewing angles, polarization is still parallel everywhere along the jet axis. For $\theta_{\rm obs}=\Theta_{\rm j}$, the polarization degree profile is less uniform, with polarization degree maxima at the level of $\sim 42\%$ around the jet origin and reconfinement point, and much lower polarization degrees in the middle region. For $\theta_{\rm obs}=2\Theta_{\rm j}$ and $\theta_{\rm obs}=30^\circ$, polarization profiles are uniform with maximum degrees of $\sim 21\%$ and $\sim 3\%$, respectively. Note that the highest sensitivity of the polarization degree on $f$ is evident for $\theta_{\rm obs}=\Theta_{\rm j}$ and $\theta_{\rm obs}=2\Theta_{\rm j}$.

\begin{figure}
\includegraphics[width=\columnwidth]{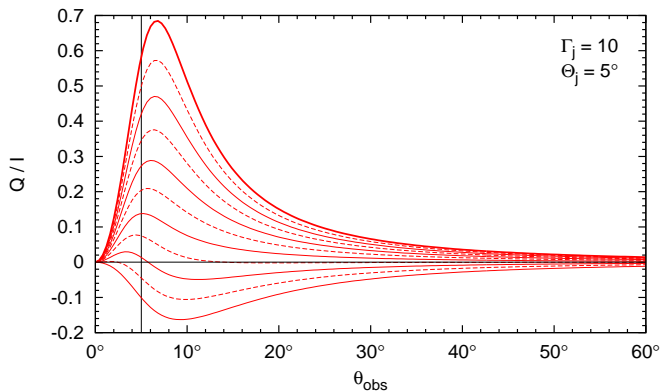}
\caption{Average polarization degree as a function of the viewing angle $\theta_{\rm obs}$ calculated for different combinations of toroidal and chaotic magnetic field components. The \emph{thick line} on the top corresponds to a purely toroidal magnetic field ($f=1$), the consecutive lines are calculated with steps $\Delta f=0.1$ down to $f=0$ corresponding to a purely chaotic magnetic field. The shock structure is the same as in the models shown in Figures \ref{fig_pol_map-1a-f100} and \ref{fig_pol_prof-pol-1a-tor}. The \emph{vertical line} indicates the jet opening angle of $\Theta_{\rm j}=5^\circ$.}
\label{fig_pol_total-tor}
\end{figure}


In Figure \ref{fig_pol_total-tor}, we show the average polarization degree as a function of the viewing angle for several values of $f$, spanning between purely toroidal magnetic fields ($f=1$) and purely chaotic magnetic fields ($f=0$). The polarization degree changes monotonically with $f$ for every viewing angle. For the purely toroidal magnetic field, the average polarization degree is always parallel. The highest polarization degree, found at $\theta_{\rm obs}\sim 7^\circ\sim 1.4\Theta_{\rm j}$, is very close to the maximum value $\Pi_{\rm max}\sim 69\%$ (for electron distribution index $p=2$). Around this viewing angle, the average polarization from a reconfinement shock depends very strongly on the value of $f$. One can obtain a wide range of polarization degrees, from $\sim 16\%$ perpendicular to $\sim 70\%$ parallel by adjusting the balance between ordered and chaotic magnetic field components. On the other hand, observation of very high polarization degrees from reconfinement shocks puts rather strong constraints on the actual viewing angle.

For small viewing angles, $\theta_{\rm obs}<3^\circ$, the perpendicular polarization signal produced by the chaotic magnetic field will be canceled by the parallel polarization from the toroidal field for $f\sim 0.1$, while for large viewing angles, $\theta_{\rm obs}>20^\circ$, this depolarization will occur at $f\sim 0.3$.

\subsection{Helical magnetic field}
\label{sec_pol_heli}

Here, we investigate the case of non-vanishing pitch angle of the magnetic field lines, $\alpha_{\rm B,m}\ne 0$. Equation (\ref{eq_pol_alphabs}) indicates that the local post-shock pitch angles at different points of the shock front will vary from $\arctan\left(\kappa\tan\alpha_{\rm B,m}\right)$ at the jet maximum radius $r_{\rm m}$ to $90^\circ$ when $r_{\rm s}\to 0$, i.e. close to the jet origin and the reconfinement point. Using analytical approximation for the shock structure \citep{2009MNRAS.392.1205N} for the flat external pressure distribution ($\eta=0$), the pitch angle can be explicitly estimated as
\be
\tan\alpha_{\rm B,s}(z) = \frac{\kappa\tan\alpha_{\rm B,m}}{4\left(z/z_{\rm r}\right)\left(1-z/z_{\rm r}\right)}\,.
\ee
In our numerical model for $\eta=0$, the shock compression parameter $\kappa$ changes from $0.25$ for $z\ll z_{\rm r}$ to $\sim 0.2$ for $z\to z_{\rm r}$. Hence, the dependence of the downstream pitch angle on $z$ is not very different from that of the upstream pitch angle.

\begin{figure}
\includegraphics[width=0.9\columnwidth]{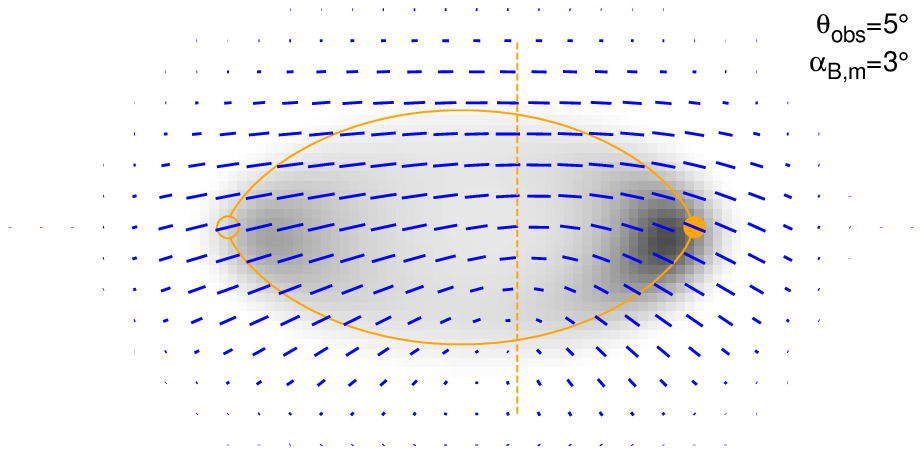}
\includegraphics[width=0.9\columnwidth]{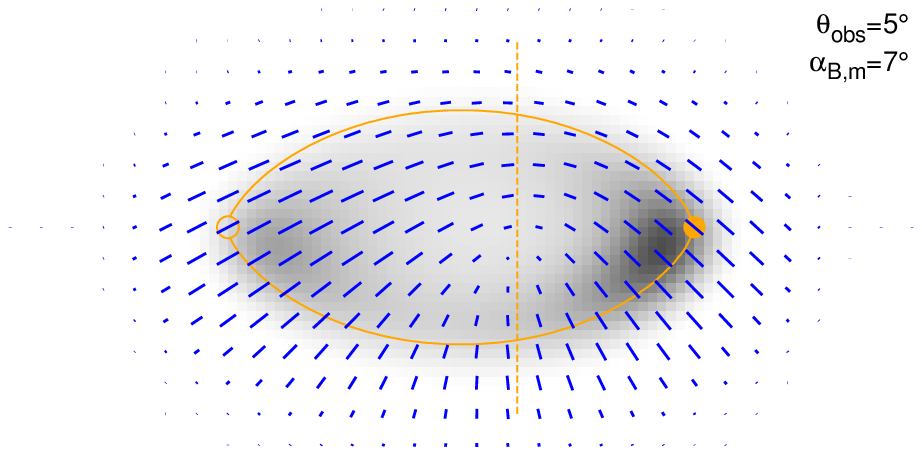}
\includegraphics[width=0.9\columnwidth]{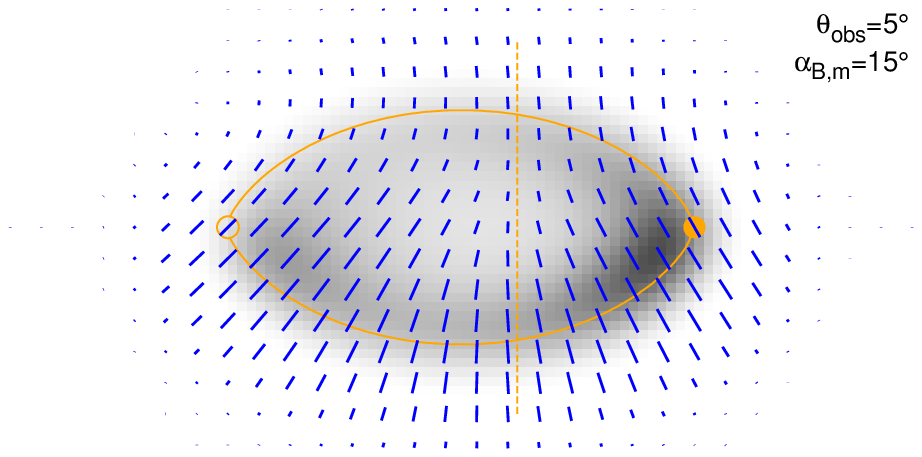}
\caption[Emission maps for helical magnetic fields]{Synthetic emission maps of the reconfinement shock with jet Lorentz factor $\Gamma_{\rm j}=10$, opening angle $\Theta_{\rm j}=5^\circ$ and external pressure index $\eta=0$, filled with a helical magnetic field of different minimum pitch angle $\alpha_{\rm B,m}$ and seen from the viewing angle $\theta_{\rm obs}=\Theta_{\rm j}$. \emph{Dashed orange lines} mark the turning point for polarization angle profiles in Figure \ref{fig_pol_prof-pol-1a-alpha}. See Figure \ref{fig_pol_map-1a-f100} for more explanation.}
\label{fig_pol_map-1a-alpha}
\end{figure}

In Figure \ref{fig_pol_map-1a-alpha}, we show emission maps for three values of $\alpha_{\rm B,m}=3^\circ, 7^\circ, 15^\circ$. They are calculated for the same shock structure as the maps calculated for the toroidal magnetic field ($\alpha_{\rm B,m}=0$) shown in Figure \ref{fig_pol_map-1a-f100}. The viewing angle is fixed at $\theta_{\rm obs}=\Theta_{\rm j}$. Our first observation is that both the total and polarized flux maps are not symmetric with respect to the projected jet axis. This is because the observed brightness depends on the orientation of the magnetic field lines, while the projection of the helical magnetic field distribution is not axisymmetric, even if it has an axial symmetry in the three-dimensional space. As a consequence of this asymmetry, the integrated Stokes parameter $U$ will not vanish. The average polarization angle will not be strictly parallel nor perpendicular to the jet axis.

Maps of the total flux have a brighter lower rim and a dimmer upper rim with increasing $\alpha_{\rm B,m}$. The bright spots are shifted toward the lower rim and become less prominent. Longitudinal profiles of the total flux (not shown) show a systematic increase of the flux from the middle region relative to the peak around the reconfinement point. For $\alpha_{\rm B,m}=15^\circ$, the bright spots become very distorted and the upper rim becomes visible. For all values of $\alpha_{\rm B,m}$, the central region of the image contributes little to the integrated flux.

Polarization vectors show a dramatic structural change with a moderate increase of the minimum pitch angle. For the toroidal magnetic field ($\alpha_{\rm B,m}=0$), the polarization is strong and parallel across the entire image. For $\alpha_{\rm B,m}=3^\circ$, the emission from the lower rim shows significantly lower polarization degrees, at one point it is almost completely depolarized. Polarization vectors corresponding to the main bright spot around the reconfinement point appear to rotate clockwise (CW) by $\sim 20^\circ$ with respect to the jet axis, while those around the secondary bright point around the jet origin appear to rotate counterclockwise (CCW) by $\sim 15^\circ$. Polarization vectors on the upper rim show little change compared to those for $\alpha_{\rm B,m}=0$.

For $\alpha_{\rm B,m}=7^\circ$, the polarization degrees decrease in the central region and along the upper rim, but remain very high around the bright spots. The polarization vectors are aligned with the upper rim and perpendicular to the lower rim. Around the main bright spot the polarization vectors are rotated CW by $\sim 40^\circ$, and around the secondary spot they are rotated CCW by $\sim 30^\circ$.

For $\alpha_{\rm B,m}=15^\circ$, polarization degrees are significantly higher along the lower rim than the upper rim. Their orientation is roughly perpendicular to the lower rim and perpendicular to the projected jet axis along the upper rim. Polarization vectors at the projected reconfinement point are rotated CW by $\sim 60^\circ$ and those at the projected jet origin are rotated CCW by $\sim 45^\circ$.

Not surprisingly, increasing the pitch angle in the helical magnetic field distribution has the effect of turning the polarization vectors from parallel to perpendicular with respect to the projected jet axis. At the same time, the total emission pattern changes from that dominated by bright spots at both ends of the reconfinement structure to that dominated by one of the rims. The average polarization degree can be expected to be very sensitive to the value of $\alpha_{\rm B,m}$.

\begin{figure}
\includegraphics[width=\columnwidth]{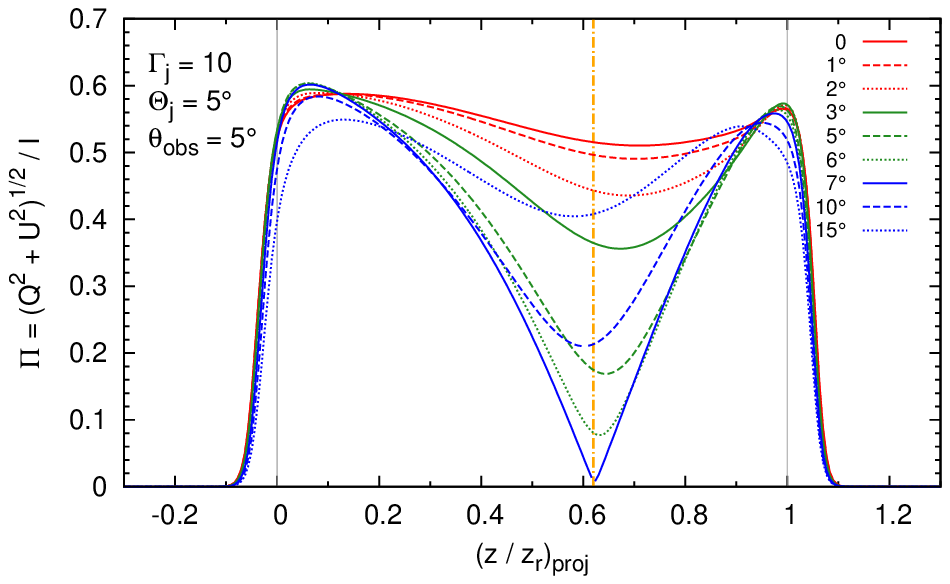}
\includegraphics[width=\columnwidth]{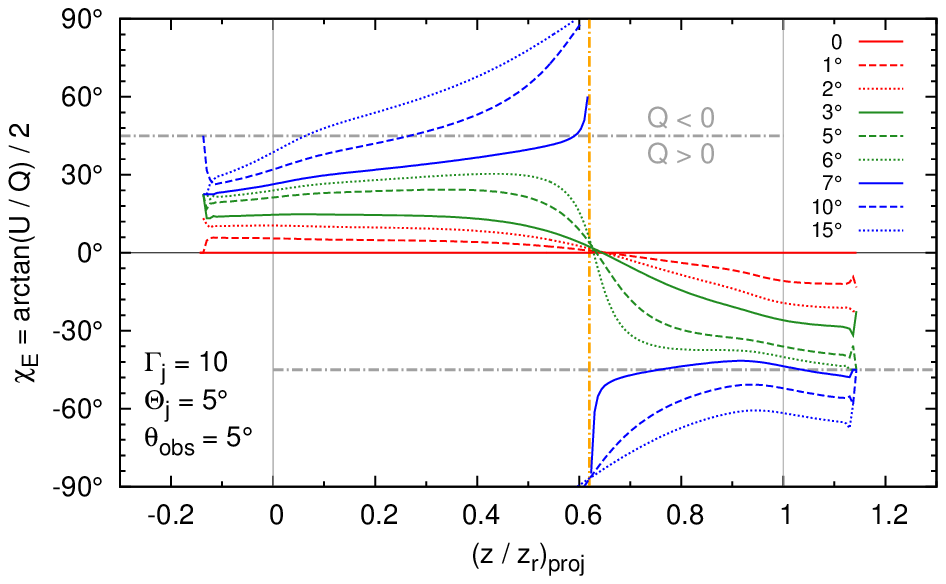}
\caption[Profiles of polarization degree and angle for helical magnetic fields]{Profiles of the average polarization degree (\emph{upper panel}) and the polarization angle (\emph{lower panel}) along the projected jet axis, calculated for a helical magnetic field with different values of the minimum pitch angle $\alpha_{\rm B,m}$, indicated in the legend. Both panels show exactly the same models, three of them are shown in Figure \ref{fig_pol_map-1a-alpha}. See Figure \ref{fig_pol_prof-pol-1a-tor} for more explanation.}
\label{fig_pol_prof-pol-1a-alpha}
\end{figure}


In Figure \ref{fig_pol_prof-pol-1a-alpha}, we show the longitudinal profiles of the polarization degree for $\alpha_{\rm B,m}$ ranging from $0$ to $15^\circ$. Since the Stokes parameter U integrated over slices of the same $z_{\rm proj}$ is non-zero, the polarization degree and angle are plotted separately. The case of $\alpha_{\rm B,m}=0$ is shown with the \emph{dashed line} in the \emph{upper panel} of Figure \ref{fig_pol_prof-pol-1a-tor}. With increasing values of the pitch angle, the polarization degree decreases in the middle part of the projected jet. At the same time, it remains very high ($\sim 55-60\%$) around the positions of the jet origin and the reconfinement point. For $\alpha_{\rm B,m}\sim 7^\circ$, the average emission becomes depolarized at $(z/z_{\rm r})_{\rm proj}\sim 0.62$. We call this point the \emph{turning point} and mark it with the \emph{vertical dashed lines} in Figures \ref{fig_pol_map-1a-alpha} and \ref{fig_pol_prof-pol-1a-alpha}. For higher values of $\alpha_{\rm B,m}$, the polarization degree increases in the middle section and slightly decreases at the extremes. For $\alpha_{\rm B,m}=15^\circ$, the polarization degree at the turning point is at the level of $\sim 40\%$.

\begin{figure}
\includegraphics[width=\columnwidth]{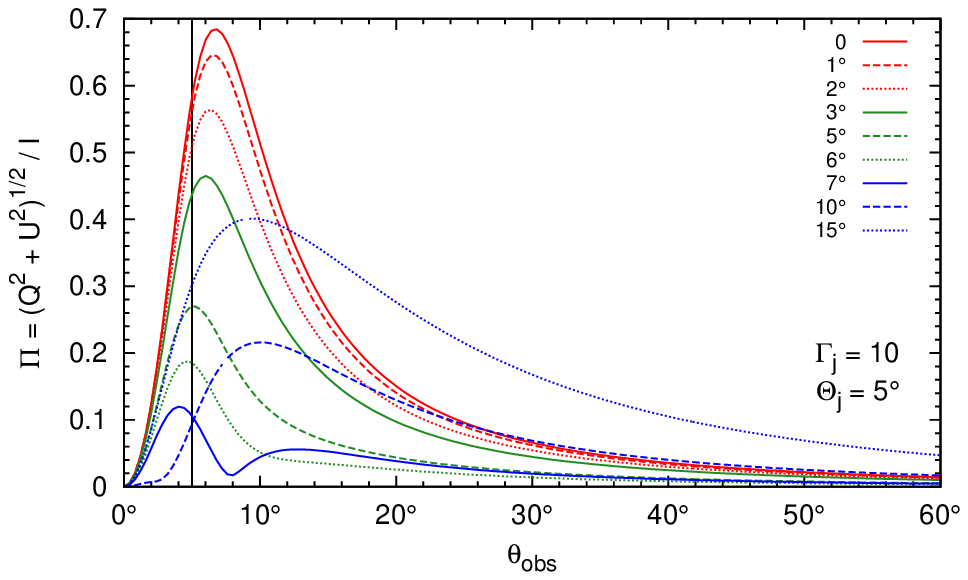}
\includegraphics[width=\columnwidth]{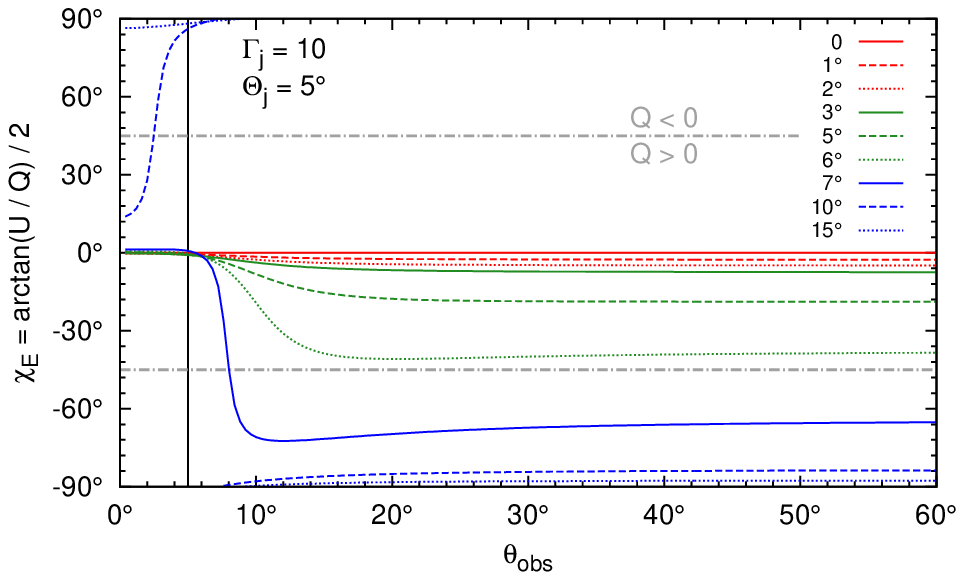}
\caption[Average polarization degree and angle for helical magnetic fields]{Average polarization degree (\emph{upper panel}) and polarization angle (\emph{lower panel}) as a function of the viewing angle, calculated for a helical magnetic field with different values of the minimum pitch angle $\alpha_{\rm B,m}$, indicated in the legend. The shock structure is the same as in the models shown in Figures \ref{fig_pol_map-1a-alpha} and \ref{fig_pol_prof-pol-1a-alpha}. \emph{Solid vertical line} indicates the jet opening angle $\Theta_{\rm j}=5^\circ$.}
\label{fig_pol_total-heli}
\end{figure}


For small pitch angles, the polarization angle increases (rotates CCW) before the turning point and decreases otherwise. For $\alpha_{\rm B,m}<7^\circ$ the modulus of the polarization angle does not exceed $45\%$, i.e. $Q>0$ and the polarization can be described as closer to parallel than perpendicular. Also, the polarization at the turning point is parallel. This changes for $\alpha_{\rm B,m}>7^\circ$, when the polarization vectors rotate mostly CCW with increasing $z_{\rm proj}$ and the polarization at the turning point becomes perpendicular. For $\alpha_{\rm B,m}=15^\circ$, the modulus of the polarization angle is generally higher than $45\%$, hence $Q<0$.

The nature of the turning point is that the average value of Stokes parameter $U$ is always close to $0$. The case of $\alpha_{\rm B,m}\sim 7^\circ$ is transitional, because the average Stokes parameter $Q$ changes sign there. It is also evident in the emission maps in Figure \ref{fig_pol_map-1a-alpha}. Polarization vectors in the vicinity of the turning point are either parallel or perpendicular to the projected jet axis. In the map for $\alpha_{\rm B,m}=3^\circ$, the polarization is mostly parallel or very low. For $\alpha_{\rm B,m}=15^\circ$, the polarization is perpendicular. And for $\alpha_{\rm B,m}=7^\circ$, roughly half of the vectors are parallel and the other half perpendicular, with relatively uniform total flux.

In Figure \ref{fig_pol_total-heli}, we show the average polarization degree and angle as functions of the viewing angle for the same values of $\alpha_{\rm B,m}$ as in Figure \ref{fig_pol_prof-pol-1a-alpha}. The average polarization degree initially decreases with the increasing pitch angle. The average polarization angle is very close to $0$ for $\theta_{\rm obs}<\Theta_{\rm j}$ and rotated in the CW direction, by a value roughly proportional to $\alpha_{\rm B,m}$, for $\theta_{\rm obs}>\Theta_{\rm j}$. It does not depend on the viewing angle for $\theta_{\rm obs}>20^\circ$.

The case of $\alpha_{\rm B,m}\sim 7^\circ$ is once again of special interest. The polarization degree for all viewing angles is lower than $\sim 11\%$. The polarization angle changes sharply between $\sim 0$ for $\theta_{\rm obs}=\Theta_{\rm j}$ to $\sim -70^\circ$ for $\theta_{\rm obs}=2\Theta_{\rm j}$ and remains at this level for larger viewing angles. For higher values of the minimum pitch angle, the polarization degree increases and the polarization angle is generally close to $90^\circ$, i.e. polarization is perpendicular to the jet axis. For $\alpha_{\rm B,m}=10^\circ$, polarization angle deviates from $90^\circ$ for very small viewing angles, but the corresponding polarization degrees are very low. For $\alpha_{\rm B,m}=15^\circ$, the maximum polarization degree is $\sim 40\%$ for $\theta_{\rm obs}\sim 9^\circ$.

\section{Discussion}
\label{sec_dis}

Polarization surveys, combined with knowledge of the inner jet structure inferred from VLBI imaging, can provide information on the orientation of magnetic fields with respect to the projected jet axis. Optical polarization electric vectors in compact radio sources are preferentially parallel to the inner jet \citep[\eg][]{1985AJ.....90...30R}. This polarization alignment can be explained with transverse internal shocks. In the reconfinement shock scenario discussed in this work, strong parallel polarization can only be produced in the presence of ordered magnetic fields that contribute at least half the total magnetic energy density (see Figure \ref{fig_pol_total-tor}) and have a minimum pitch angle of less than $\sim 5^\circ$ (see Figure \ref{fig_pol_total-heli}).

Polarization measured in large-scale jets between prominent knots is usually perpendicular, both in radio \citep[\eg][]{1994AJ....108..766B} and optical bands \citep[\eg][]{2006ApJ...651..735P}. It indicates the presence of a strong poloidal component of the magnetic field. However, in expanding jet the poloidal magnetic field component decays faster than the toroidal component. In large-scale jets, the poloidal component of the magnetic field can arise through the velocity shear at the jet boundary \citep{1981ApJ...248...87L}. \cite{2006MNRAS.367..851C} studied polarization of synchrotron radiation from conical shocks filled with a combination of chaotic and poloidal magnetic fields. He calculated polarized emission maps and compared them with resolved polarimetric VLBI map of blazar 3C~380. He was able to reproduce a fan-like structure of the polarization vectors for an equal contribution of chaotic and poloidal magnetic field components ($f=0.5$). His motivation for introducing the poloidal component was to obtain higher degrees of the perpendicular polarization. Reconfinement shocks or conical shocks with diverging upstream flow can produce higher perpendicular polarization degrees without an ordered magnetic field component. 

\cite{2005MNRAS.360..869L} studied polarization from helical magnetic fields in relativistic cylindrical jets. In this case, the pitch angle is constant along the jet and therefore their results are difficult to compare directly with the results of Section \ref{sec_pol_heli}, where the pitch angle is strongly position-dependent. However, the dependence of the average polarization degree on the viewing angle and the minimum pitch angle (Figure \ref{fig_pol_total-heli}) is in qualitative agreement with their Figure 7a. One important difference is that in their model the average Stokes parameter $\left<U\right>$ vanishes. In our model, $\left<U\right>\ne 0$, since we have defined the poloidal component of the magnetic field along the fluid velocity unit vector $\bm{e}$ (see Equation \ref{eq_pol_vecb_heli}). Only when $\bm{e}$ is parallel to the jet axis, as is the case for a cylindrical jet model, $\left<U\right>=0$. This provides a physical interpretation for the turning point found in Figure \ref{fig_pol_prof-pol-1a-alpha}. It corresponds to a distance $z$ along the jet, for which the post-shock velocity inclination angle $\theta_{\rm s}=0$. Since $\theta_{\rm s}$ is a monotonically decreasing function of $z$, there can be only one such turning point, located closely to the point of maximum jet radius.

\cite{2005MNRAS.360..869L} also studied the transverse profiles of polarization degree. In their Figure 9a,b they show that those profiles should be symmetric with respect to the jet axis for $\theta_{\rm obs}=1/\Gamma$ and strongly asymmetric for $\theta_{\rm obs}=1/(2\Gamma)$ or $\theta_{\rm obs}=2/\Gamma$. Polarization vectors close to the jet boundary should always be perpendicular. Using the emission maps shown in Figure \ref{fig_pol_map-1a-alpha}, one can evaluate the transverse polarization degree profiles along the lines indicating the turning point. These models correspond to $\theta_{\rm obs}\Gamma_{\rm j}=0.9$ and $\theta_{\rm obs}\Gamma_{\rm s}$ slightly lower. The $Q/I$ profiles are strongly asymmetric, with the $Q/I$ value increasing from the lower rim to the upper rim. Polarization vectors are not always perpendicular at the jet boundary, but this may be due to the low resolution of the maps. Our polarization profiles appear to be roughly consistent with their case of $\theta_{\rm obs}=1/(2\Gamma)$. This may indicate that the transverse polarization profile for cylindrical jets filled with helical magnetic fields is very sensitive to $\theta_{\rm obs}$ for values close to $1/\Gamma$.

On the other hand, some observed radio structures in jets can be understood without invoking any ordered magnetic field component. An interesting recent example is the C80 knot in the jet of radio galaxy 3C~120, reported by \cite{2012ApJ...752...92A}. It has a bow-like shape with polarization vectors aligned perpendicularly to its outline. It was well reproduced by a conical shock model with compressed chaotic magnetic field distribution. However, this model can hardly explain the polarization vectors perpendicular to the jet axis observed immediately downstream of the C80 knot, even when a poloidal field component is introduced. Additional perpendicular shock waves can explain a parallel polarization measured even farther downstream (knot C99), but are inconsistent with a perpendicular polarization. We propose a simple explanation of the perpendicular polarization vectors downstream of the C80 knot by a gradual collimation of the conical shock into a roughly cylindrical structure. In the absence of large-scale magnetic fields, the polarization vectors will be perpendicular to the local shock outline \citepalias{2009MNRAS.395..524N}.

Our model predicts a relatively uniform distribution of the total brightness, especially at large viewing angles. Hence, it cannot reproduce very bright and compact features like HST-1 in radio galaxy M87. Clearly, an additional dissipation mechanism is required to explain its behavior, in particular the high optical/UV polarization reported by \cite{2011ApJ...743..119P}.

\section{Summary}
\label{sec_sum}

We calculated emission maps and longitudinal profiles of total flux and polarization degree, and average polarization degree for relativistic reconfinement shocks with different combinations of chaotic and ordered magnetic field distributions.

For toroidal magnetic fields (Section \ref{sec_pol_tor}), the average polarization is always parallel to the jet axis, and the average polarization degree can reach the maximum value allowed for a particular electron distribution. The total flux emission maps for $\theta_{\rm obs}>\Theta_{\rm j}$ are similar to those for the case of chaotic magnetic fields, for small viewing angles they lack emission from the rims.

For helical magnetic fields (Section \ref{sec_pol_heli}), which are characterized by the value of the minimum pitch angle $\alpha_{\rm B,m}$ measured at the point of maximum jet radius, emission maps are asymmetric with respect to the projected jet axis and hence the Stokes parameter $U\ne 0$. For $\alpha_{\rm B,m}\gtrsim 7^\circ$, the total flux is concentrated on one side of the jet axis, where polarization vectors align with the shock outline and the polarization degrees are higher than on the other side. Longitudinal profiles show a characteristic turning point at $(z/z_{\rm r})_{\rm proj}\sim 0.6$, where average $U\sim 0$, regardless of the pitch angle. It corresponds to the point, at which the post-shock velocity vectors are parallel to the jet axis, as for the cylindrical jet models studied by \cite{2005MNRAS.360..869L}. It separates the initial jet section, in which polarization vectors are rotated counterclockwise to the jet axis from the section in which the polarization vectors are rotated clockwise. The average polarization angle is $|\chi_{\rm E}|<45^\circ$ ($Q>0$) for $\alpha_{\rm B,m}<7^\circ$, and the opposite is generally true (see Fig. \ref{fig_pol_total-heli}). For $\alpha_{\rm B,m}\sim 7^\circ$, the reconfinement shock appears to be effectively depolarized for all observers.

\acknowledgements{This work has been partly supported by the Polish MNiSW grants N~N203 301635 and N~N203 386337, the Polish ASTRONET grant 621/E-78/SN-0068/2007, the Polish NCN grant DEC-2011/01/B/ST9/04845, and the NSF grant AST-0907872 and the NASA Astrophysics Theory Program grant NNX09AG02G.}

\appendix

\section{Polarization and anisotropy of the synchrotron radiation}
\label{app_pol}

In Section \ref{sec_pol}, we calculated the polarization properties of synchrotron radiation arising from different distributions of the magnetic field. The total emitted power is constrained, but the intrinsic anisotropy of the synchrotron emission has to be accounted for. The exact results depend on the index $p$ of the electron energy distribution $N(\gamma)\propto\gamma^{-p}$. However, \cite{1980MNRAS.193..439L} and \cite{1985ApJ...298..301H} showed that the formulas are greatly simplified for $p=3$.

\subsection{Uniform magnetic field}

For a uniform magnetic field $\bm{B}$, we have
\bea
I(\bm{k}) &=& CB_\perp^2\,, \\
Q(\bm{k}) &=& I(\bm{k})\Pi_{\rm max}\cos{(2\chi_{\rm E})}\,, \\
U(\bm{k}) &=& I(\bm{k})\Pi_{\rm max}\sin{(2\chi_{\rm E})}\,,
\eea
where $\bm{k}$ is the unit vector toward the observer, $B_\perp$ is the magnetic field component normal to the line of sight, $\chi_{\rm E}$ is the electric vector polarization angle, $\Pi_{\rm max}=(3p+3)/(3p+7)$ is the maximum polarization degree of the synchrotron radiation, and $C$ is a constant. We note that $B_\perp^2=B^2-(\bm{B}\cdot\bm{k})^2$ and $\chi_{\rm E}=\chi_{\rm B}-\pi/2$, where $\chi_{\rm B}$ is the magnetic field polarization angle. We introduce orthogonal coordinates in the plane of the sky $(\bm{v},\bm{w})$, in which the polarization angle is measured. The magnetic vector polarization angle is
\be
\tan\chi_{\rm B}=\frac{\bm{B}\cdot\bm{w}}{\bm{B}\cdot\bm{v}}\,,
\ee
hence
\bea
\cos{(2\chi_{\rm E})} = \frac{\tan^2\chi_{\rm B}-1}{\tan^2\chi_{\rm B}+1} &=& \frac{(\bm{B}\cdot\bm{w})^2-(\bm{B}\cdot\bm{v})^2}{B_\perp^2}\,, \\
\sin{(2\chi_{\rm E})} = \frac{-2\tan\chi_{\rm B}}{\tan^2\chi_{\rm B}+1} &=& \frac{-2(\bm{B}\cdot\bm{v})(\bm{B}\cdot\bm{w})}{B_\perp^2}\,.
\eea
General formulas for the Stokes parameters are
\bea
I(\bm{k}) &=& C\left[B^2-(\bm{B}\cdot\bm{k})^2\right]\,, \\
Q(\bm{k}) &=& C\Pi_{\rm max}\left[(\bm{B}\cdot\bm{w})^2-(\bm{B}\cdot\bm{v})^2\right]\,, \\
U(\bm{k}) &=& C\Pi_{\rm max}\left[-2(\bm{B}\cdot\bm{v})(\bm{B}\cdot\bm{w})\right]\,.
\eea

The $C$ constant can be found from the total radiation flux:
\be
\Phi=\int d\Omega_{\bm{k}}I(\bm{k})=\frac{8\pi}{3}CB^2
\quad
\Rightarrow
\quad
C=\frac{3\Phi}{8\pi B^2}\,.
\ee

\subsection{Compressed chaotic magnetic field}

Following the calculations of \cite{1985ApJ...298..301H}, we consider an isotropic magnetic field distribution
\be
\bm{B}=B\bm{b}=B[\sin\theta\cos\phi,\sin\theta\sin\phi,\cos\theta]\,.
\ee
It is then compressed along unit vector $\bm{n}=[0,0,1]$ by factor $\kappa\le 1$
\be
\tilde{\bm{B}}=B\left[\frac{1}{\kappa}\sin\theta\cos\phi,\frac{1}{\kappa}\sin\theta\sin\phi,\cos\theta\right]\,.
\ee
The average Stokes parameters are found by integration over all possible unit vectors $\bm{b}$. For an observer located at $\bm{k}=[\sin\alpha,0,\cos\alpha]$ (and plane-of-the-sky coordinate vectors $\bm{v}=[\cos\alpha,0,-\sin\alpha]$ and $\bm{w}=[0,1,0]$), the results are
\bea
I(\bm{k}) &=& \frac{C}{4\pi}\int d\Omega_{\bm{b}}\left[\tilde{B}^2-(\tilde{\bm{B}}\cdot\bm{k})^2\right] =\nonumber\\
&=&\frac{CB^2}{3\kappa^2}\left[2-\left(1-\kappa^2\right)\sin^2\alpha\right]\,, \\
Q(\bm{k}) &=& \frac{C\Pi_{\rm max}}{4\pi}\int d\Omega_{\bm{b}}\left[(\tilde{\bm{B}}\cdot\bm{w})^2-(\tilde{\bm{B}}\cdot\bm{v})^2\right] = \nonumber\\ &=& \frac{C\Pi_{\rm max}B^2}{3\kappa^2}\left(1-\kappa^2\right)\sin^2\alpha\,, \\
U(\bm{k}) &=& \frac{C\Pi_{\rm max}}{4\pi}\int d\Omega_{\bm{b}}\left[-2(\tilde{\bm{B}}\cdot\bm{v})(\tilde{\bm{B}}\cdot\bm{w})\right] = 0\,.
\eea
Hence, the polarization degree is
\be
\label{eq_app_polar_pd}
\Pi(\bm{k}) = \frac{Q(\bm{k})}{I(\bm{k})} = \Pi_{\rm max}\frac{\left(1-\kappa^2\right)\sin^2\alpha}{2-\left(1-\kappa^2\right)\sin^2\alpha}\,,
\ee
in agreement with \cite{1985ApJ...298..301H}. The polarization angle is $\chi_{\rm E}=0$, since $Q(\bm{k})>0$ and $U(\bm{k})=0$. The $C$ constant is found from
\be
\Phi = \int d\Omega_{\bm{k}}I(\bm{k}) = \frac{8\pi\left(2+\kappa^2\right)}{9\kappa^2}CB^2\,,
\ee
hence
\be
C = \frac{9\kappa^2\Phi}{8\pi\left(2+\kappa^2\right)B^2}\,.
\ee

\end{document}